\documentclass[sigconf]{acmart}
\usepackage{tabularx}
\AtBeginDocument{%
  }

\setcopyright{none}
\renewcommand\footnotetextcopyrightpermission[1]{}

\usepackage{float}

\begin{document}

\title{\textit{AuthScribe}: Supporting Agency and Authorship in LLM-Assisted Reflective Journaling}

\author{Lanping Hu}
\affiliation{%
  \institution{Australian National University}
  \country{Australia}
}

\author{Yinsi Zhou}
\affiliation{%
  \institution{University of New South Wales}
  \country{Australia}
}

\author{Zhenchang Xing}
\affiliation{%
  \institution{CSIRO Technology }
  \country{Australia}
}

\author{Xiaoyu Sun}
\affiliation{%
  \institution{Australian National University}
  \country{Australia}
}

\author{Shidong Pan}
\affiliation{%
  \institution{New York University}
  \country{USA}
}

\renewcommand{\shortauthors}{Hu et al.}

\begin{abstract}
Large language models are increasingly used in reflective journaling, where articulating an experience is part of interpreting it. This raises questions about how writers retain authorship, exercise agency, and assess personalized assistance. We present AuthScribe, an LLM-assisted reflective journaling system that combines contextual anchors, writer-steered AI support, Autobiographical Citations, and read-only Provisional Drafts that require re-expression before AI-generated suggestions enter the journal. Retrieval and LLM inference run on-device. In a 14-day field deployment with 14 participants, grounding shifted from situating current experiences to reconnecting them with prior self-authored interpretations. Perceived agency remained stable through Week 1 and shifted during Week 2. Participants described authorship through transforming AI-generated suggestions into personally meaningful expression, and trusted personalized prompts when they could trace them to prior writing. We discuss design implications for adoption friction, longitudinal agency, and traceable personalization in AI systems for personal reflection.
\end{abstract}

\ccsdesc[500]{Human-centered computing~Empirical studies in HCI}
\ccsdesc[300]{Human-centered computing~Interaction design theory, concepts and paradigms}


\keywords{reflective journaling, human-AI interaction, authorship, sense of agency, personalization, adoption friction}

\maketitle

\section{Introduction}

People use reflective journaling to make sense of lived experience: they select what happened, interpret why it mattered, and carry that interpretation into an evolving account of themselves. Research on expressive writing and narrative identity connects such practices to coherent personal narratives \cite{pennebaker1999forming,Habermas_2000}. HCI research describes reflection as active inquiry in which people revisit evidence, question existing interpretations, and develop new understandings \cite{baumer2015reflective,Bentvelzen_2022,cho2022reflection}. For journal writers, wording is part of this work. Choosing an episode, giving it meaning, and committing an interpretation to language shape the reflection itself.

LLM-assisted reflective journaling systems increasingly participate in these activities. Recent systems help people begin writing, elaborate daily experiences, explore personal challenges, and connect reflection to behavioral traces or prior self-authored material \cite{mindfuldiary,diarymate,nepal2024mindscape,Zhou_2025,song2025exploreself}. Memory-oriented systems bring prior experiences and accumulated records into present reflection \cite{zulfikar2025resonance,ozawaburns2025remigraph,hannan2025ltj}. This work expands AI support from articulating experience to interpreting it. A particularly close example is Augmentiary, which presents candidate interpretations of journal passages for writers to compare with their own understanding over a four-week deployment \cite{hwang2026augmentiary}. This interpretive capability gives writers useful material for reflection and places AI-generated meaning closer to the construction of a personal narrative.

For writers, this shift creates a consequential question of authorship. Human--AI writing research shows that AI-generated suggestions can influence what people express, which ideas they pursue, how much cognitive effort they invest, and how much ownership they experience over the resulting text \cite{jakesch2023opinionated,Draxler_2024,pan2024don,lee2025critical,bhat2026reactive,park2026authorship}. Studies across writing contexts also show that users value opportunities to exercise judgment over consequential parts of the writing process, and that interaction structure shapes where they experience agency and ownership \cite{nishal2026agency,carrera2026egg,polimetla2026ownership,Reza_2025}. The AI Ghostwriter Effect captures one version of this tension: writers can identify themselves as authors while experiencing weaker ownership over AI-generated passages \cite{Draxler_2024}. In reflective journaling, an AI continuation can carry a ready-made interpretation of a private event. The writer decides how that interpretation enters an account they are willing to claim as their own.

Personalization creates a second consequential relationship. AI systems for personal reflection increasingly draw on sensor traces, notes, prior entries, and autobiographical memories to generate personalized assistance \cite{nepal2024mindscape,zulfikar2025resonance,ozawaburns2025remigraph,lin2026mobile}. Personalized assistance can shape engagement, autonomy, reliance, and perceived ownership \cite{qin2026personalization,kim2025reliance}. Research on transparent and controllable AI interfaces shows the value of exposing sources, intermediate states, and actionable control points that help users evaluate system output \cite{kim2025reliance,wu2022aichains,amershi2019guidelines,rismani2026oversight}. For journal writers, the relevant question extends beyond whether a prompt feels personal. They also need a basis for understanding why the system connects a present concern to a particular part of their history.

These developments make two moments especially important in LLM-assisted reflective journaling. First, \emph{personal context enters AI assistance}: a system selects past material and turns it into a prompt, interpretation, or suggestion. Second, \emph{AI assistance enters personal writing}: the writer decides how an AI-generated suggestion enters a reflective account. Existing journaling systems provide rich support for prompting, conversational guidance, user-directed exploration, sensing, and memory retrieval \cite{mindfuldiary,diarymate,nepal2024mindscape,song2025exploreself,zulfikar2025resonance}. Human--AI writing research provides complementary evidence about ownership, critical engagement, and the influence of AI-generated suggestions \cite{Draxler_2024,Reza_2025,bhat2026reactive,park2026authorship}. Our work examines how writers experience both transitions during longitudinal use.

We explore these transitions through \textit{AuthScribe}, an LLM-assisted reflective journaling system implemented as an Obsidian plugin. AuthScribe combines contextual anchors, user-steered Reflect and Expand actions, Autobiographical Citations, and read-only Provisional Drafts. Autobiographical Citations expose the prior self-authored material used to ground a reflective prompt. Provisional Drafts keep generated continuations visually separate from editable journal text and ask writers to re-express ideas they choose to carry forward. We use \emph{adoption friction} to describe this deliberate interaction effort at the uptake boundary. During the study, AuthScribe ran retrieval and LLM inference on-device. Participants verified this during onboarding by turning off network connectivity and generating a test response.

We studied AuthScribe through a 14-day in-the-wild deployment with 14 participants. We collected non-semantic interaction metadata, administered the Sense of Agency Scale at baseline, Week~1, and Week~2, measured system usability at the end of the study, and conducted semi-structured exit interviews. Journal content remained on participants' devices throughout the evaluation. We asked:

\begin{description}
    \item[RQ1.] How do people appropriate AI-generated suggestions into reflective journaling when those suggestions remain separate from the editable journal and uptake requires re-expression?
    \item[RQ2.] How do participants' perceived agency, experiences of grounding, and trust evolve as interaction practices and personal context accumulate over longitudinal use?
\end{description}

Our findings show that participants' relationship with AI assistance changed as they developed a practice around the system. Early contextual grounding helped them convert broad self-judgments into situated questions about specific episodes; accumulated journal entries later gave them self-authored evidence for connecting interpretations across time. Participants' broad perceived agency remained stable through Week~1 and shifted during Week~2, after they had established repeated patterns of steering, rewriting, and grounding in prior self-authored material. Provisional Drafts revealed authorship as a process of transformation and commitment: participants appropriated AI-generated suggestions by changing their wording, specificity, and personal meaning before placing them in the journal. Autobiographical Citations made personalized assistance more trustworthy when participants could trace a current prompt to recognizable prior reasoning. These findings show how interaction boundaries can shape authorship, agency, and trust by making AI contributions inspectable and actionable.

This work makes three contributions:
\begin{enumerate}
    \item \textbf{AuthScribe}, an LLM-assisted reflective journaling system that makes two consequential transitions visible: Autobiographical Citations expose how prior self-authored material enters AI assistance, and Provisional Drafts structure how AI-generated suggestions enter personal writing.
    \item \textbf{A longitudinal empirical account of LLM-assisted reflective journaling}, showing how grounding changed what participants used as evidence about themselves, how perceived agency shifted after an established period of use, how authorship emerged through transformation and commitment, and how autobiographical provenance shaped trust.
    \item \textbf{Design knowledge for AI systems that support personal reflection}, including adoption friction as a lens for designing the uptake of AI-generated suggestions, accountable autobiographical provenance as a basis for inspectable personalization, and a longitudinal view of agency as a practice of use across repeated interaction.
\end{enumerate}

\section{Related Work}
\label{sec:related-work}

\subsection{Technology-Mediated Reflection and LLM-Assisted Reflective Journaling}

HCI has approached reflection as an active process through which people examine experience, question existing interpretations, and develop new understandings \cite{baumer2015reflective,Bentvelzen_2022,cho2022reflection}. Technologies for personal reflection support this work through prompts, records of prior activity, autobiographical archives, and structured opportunities to revisit experience \cite{isaacs2013echoes,Elsden_2016}. Longitudinal journaling research shows that sustained reflective practice can support exploration, organization, self-understanding, and initiative across life transitions \cite{hannan2025ltj}. Across this literature, reflective technologies shape the evidence people can revisit and the forms of interpretation they can pursue.

LLMs extend this design space by generating language that participates directly in reflection. Recent reflective journaling systems use conversational prompts, sentence suggestions, adaptive questions, behavioral sensing, and user-directed exploration to help people articulate experience and examine it from multiple perspectives \cite{mindfuldiary,diarymate,nepal2024mindscape,Zhou_2025,song2025exploreself}. Memory-oriented systems use prior writing and autobiographical records to connect present concerns with earlier experiences or imagined futures \cite{zulfikar2025resonance,ozawaburns2025remigraph}. These systems collectively move AI support toward richer forms of interpretation and longitudinal context.

Augmentiary is especially close to our research question because it treats AI-generated interpretation itself as reflective material \cite{hwang2026augmentiary}. Writers first record an experience, then compare their own understanding with candidate interpretations produced by the system. Its four-week deployment shows how interpretive feedback can connect fragmented experiences and support self-understanding while creating tensions around agency and voice. This work establishes the value of generated interpretations during reflection. AuthScribe examines what happens at the subsequent uptake boundary, where a useful generated idea can enter the writer's own journal and acquire personal authorship through re-expression.

A parallel body of journaling research emphasizes user direction over the reflective process. Users choose topics, perspectives, questions, or paths of inquiry, and adaptive AI support follows the direction of their exploration \cite{diarymate,song2025exploreself,mindfuldiary}. This work establishes steering as an important source of control during generation. AuthScribe extends the locus of control to uptake by separating generated continuations from editable journal text. Our study examines two forms of participation within the same reflective practice: directing AI assistance and appropriating its output.

\subsection{Agency, Authorship, and Appropriation in AI-Assisted Writing}

Human--AI writing research increasingly treats authorship as an interactional outcome shaped by how users participate in producing and transforming text. Suggestion format, interaction timing, editing opportunities, amount of AI contribution, and interface structure can influence perceived control, ownership, and responsibility \cite{Lee_2022,Lehmann_2022,Draxler_2024,kadoma2024ownership,Reza_2025,polimetla2026ownership}. The AI Ghostwriter Effect captures a particularly relevant tension in this literature: people can declare authorship of AI-assisted text while experiencing weaker ownership over passages produced by the model \cite{Draxler_2024}. For reflective journaling, this tension reaches both the interpretation and the wording of the final text.

A related body of evidence shows that AI-generated suggestions can reorganize the cognitive work of writing. These suggestions can seed opinions and directions that users later elaborate, while high reliance on generative AI can shift effort toward evaluation, verification, and integration \cite{jakesch2023opinionated,bhat2026reactive,lee2025critical}. These patterns describe a move from originating every element of the text toward judging and developing AI-generated suggestions. This shift makes appropriation especially important: users still make consequential decisions about which ideas deserve attention, how those ideas fit their goals, and what form they take in the final artifact.

Agency within AI-assisted writing also develops through repeated interaction. Across recent studies, users' self-efficacy, trust, prompting strategies, expectations of control, and perceived ownership change with experience and with the role assigned to AI \cite{park2026authorship,nishal2026agency,carrera2026egg,andersson2026tombwriter,qin2026personalization}. These studies converge on the importance of preserving meaningful human judgment at moments that carry personal, professional, or creative significance. For AuthScribe, the consequential moment occurs when a generated interpretation moves toward the writer's personal narrative.

HCI research on human--AI control provides a broader basis for designing these moments. Transparent interaction structures, explicit control points, and opportunities to revise system behavior can support user understanding and intervention \cite{amershi2019guidelines,wu2022aichains}. Recent work on task-specific selective friction shows that carefully placed interaction effort can encourage deliberation around AI-supported completion \cite{arun2026friction}. We build on this direction by locating friction specifically at uptake. We call this \emph{adoption friction}: interaction effort attached to the moment when users interpret, transform, and commit AI-generated suggestions to a human-authored artifact.

\subsection{Longitudinal Personalization, Provenance, and Privacy}

AI systems for personal reflection increasingly depend on personal context that develops through use. Behavioral sensing can ground prompts in recent activity, while memory-oriented systems use prior writing and autobiographical records to reconnect experiences across time \cite{nepal2024mindscape,zulfikar2025resonance,ozawaburns2025remigraph}. Earlier smart-journaling research likewise shows that people experience personal records as meaningful histories whose value emerges through revisiting and interpretation \cite{Elsden_2016}. Personalization changes as the user's record grows and new forms of personal evidence become available.

The value of personalization also depends on how users evaluate its basis. Research on human--AI reliance shows that visible sources can influence how people assess LLM responses, while transparent interaction structures and accurate mental models can shape oversight and control \cite{kim2025reliance,wu2022aichains,rismani2026oversight}. These findings position provenance as an interaction resource: source information helps users judge the evidence behind an AI intervention and its relevance to the current task.

This concern becomes especially important when the evidence is autobiographical. A personalized prompt can refer to a behavior, a past event, or an interpretation the user previously wrote about themselves. In AuthScribe, Autobiographical Citations expose this relationship by linking reflective questions to prior journal entries and Kanban notes. Our study examines how participants use that visible connection to understand why a prompt relates to them. The resulting design question concerns \emph{accountable autobiographical provenance}: how AI systems for personal reflection can make their use of a person's history inspectable enough for that person to evaluate the connection.

Journaling also makes the handling of personal data part of the user experience. Work on smart journaling and data visibility shows that people's awareness of access and observation can shape what they record and disclose \cite{Elsden_2016,saha2024observer}. Human--AI design guidelines emphasize clear communication about system capabilities and behavior \cite{amershi2019guidelines}. AuthScribe combines autobiographical provenance with on-device retrieval and inference during the study deployment, and participants verified local processing during onboarding. We examine provenance and local processing as two ways users can understand how their personal material participates in AI assistance.

The literature highlights three forms of user participation: directing reflective exploration, judging AI-generated suggestions, and evaluating how personal context enters AI assistance. AuthScribe studies all three in a longitudinal setting. Citations expose how autobiographical evidence enters AI assistance, while Provisional Drafts structure how AI-generated suggestions enter personal writing. Our study focuses on how these visible boundaries shape authorship, agency, and trust as interaction practices and personal context develop over time.

\section{AuthScribe}

%
%
\begin{figure*}[t]
  \centering
  \includegraphics[width=\textwidth]{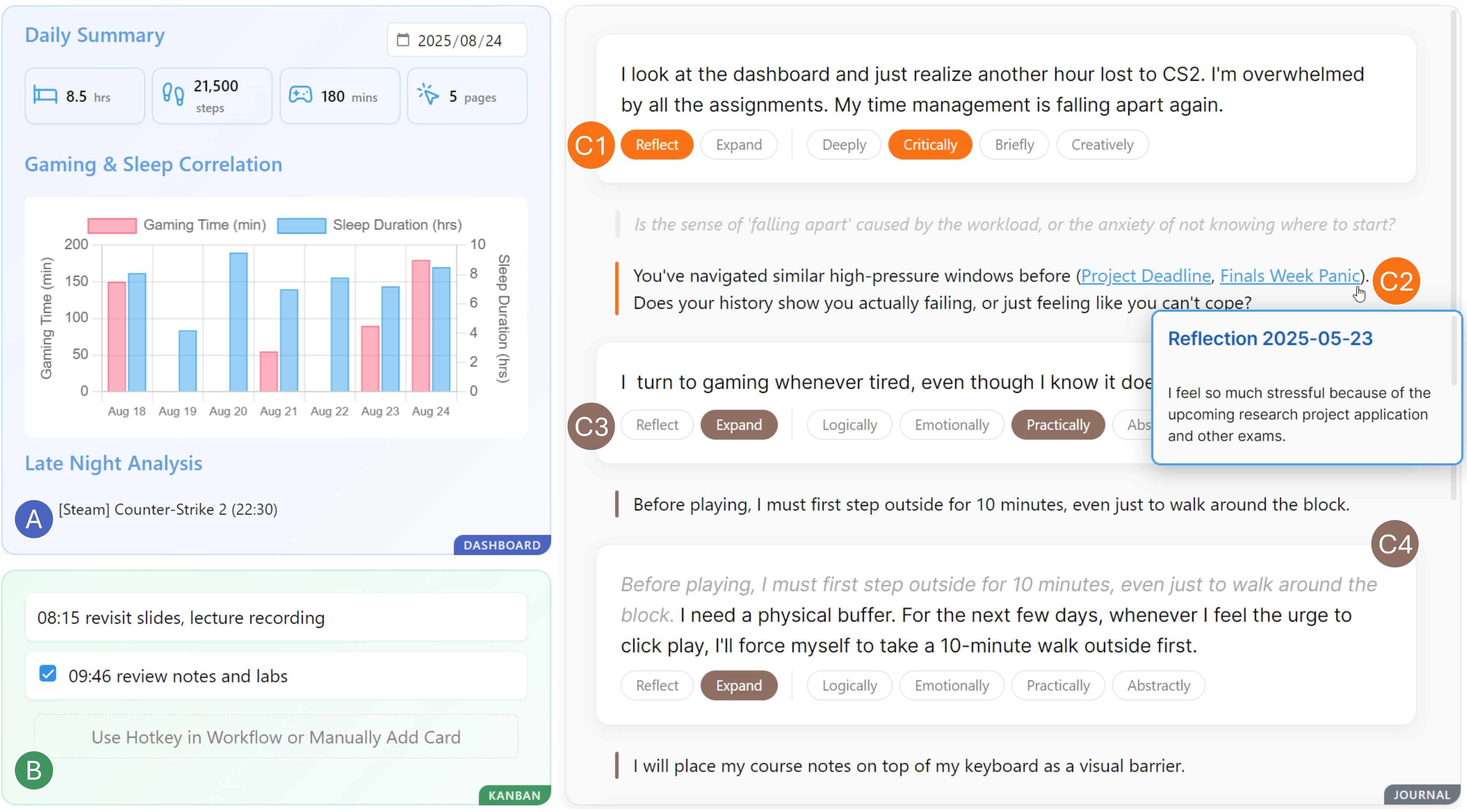}
  \caption[The active reflective journaling process using AuthScribe]{The active reflective journaling process in AuthScribe. (A) The Dashboard and (B) Kanban provide contextual anchors for reflection. (C1) \textsf{Reflect} and (C3) \textsf{Expand} let writers request AI support with an Interaction Stance. (C2) Autobiographical \textbf{Citations} connect grounded prompts to earlier personal writing. (C4) \textbf{Provisional Drafts} display generated continuations as visually distinct, read-only material that writers can re-express in the editable journal.}
  \Description{AuthScribe interface with a Dashboard and Kanban panel on the left and a reflective journaling stream on the right. The stream contains writer-authored text, Reflect and Expand controls, Autobiographical Citations, and visually distinct Provisional Drafts.}
  \label{fig:authscribe-interface}
\end{figure*}
%
%

AuthScribe is an LLM-assisted reflective journaling system implemented as an Obsidian plugin. It supports writers as they move from recalling an experience, to examining it with AI assistance, to deciding which AI-generated suggestions belong in their own reflective account. The interface combines contextual anchors, writer-steered AI actions, Autobiographical Citations, and Provisional Drafts. During the study, AuthScribe performed retrieval and LLM inference on-device. We designed AuthScribe for moments when a writer wants to examine a personal experience and needs support recalling context, developing an interpretation, or moving the reflection forward.

\subsection{Design Goals}

We summarize AuthScribe's design rationale through four goals informed by prior work on reflective journaling, user agency in AI-supported reflection, authorship in human--AI writing, and personalized reflective systems.

\paragraph{DG1: Help writers anchor reflection in specific lived experiences.}
Reflection begins by deciding which experience deserves attention and how to understand it. Daily experiences can initially appear as diffuse feelings or broad judgments about oneself. AuthScribe aims to give writers concrete evidence from their own day that they can inspect and interpret. The writer decides what those traces mean and which experiences deserve further reflection.

\paragraph{DG2: Let writers direct when and how AI participates in reflection.}
Reflective exploration follows the writer's own priorities, and those priorities can shift as an entry develops. AuthScribe gives writers control over when AI assistance enters the process and which perspective guides each intervention. This goal supports writer-led reflection while allowing AI to provide questions or possible directions when the writer chooses to engage it.

\paragraph{DG3: Keep writers active in turning AI-generated suggestions into personal expression.}
AI-generated language can introduce a complete interpretation before the writer has fully formulated one. AuthScribe aims to keep the transition from generated suggestion to personal writing active. Writers evaluate the meaning of a suggestion, decide which parts fit their experience, and determine how those ideas should appear in their own journal. This goal treats appropriation as part of authorship.

\paragraph{DG4: Make personalized assistance traceable to the writer's own history.}
Personalized prompts can draw on experiences that span multiple days and entries. Writers need a visible basis for understanding why a particular part of their history informs a current prompt. AuthScribe exposes the autobiographical evidence behind grounded assistance so writers can inspect the source, recognize the earlier context, and judge the relevance of the connection.

%
%
\begin{figure*}[t]
  \centering
  \includegraphics[width=0.8\textwidth]{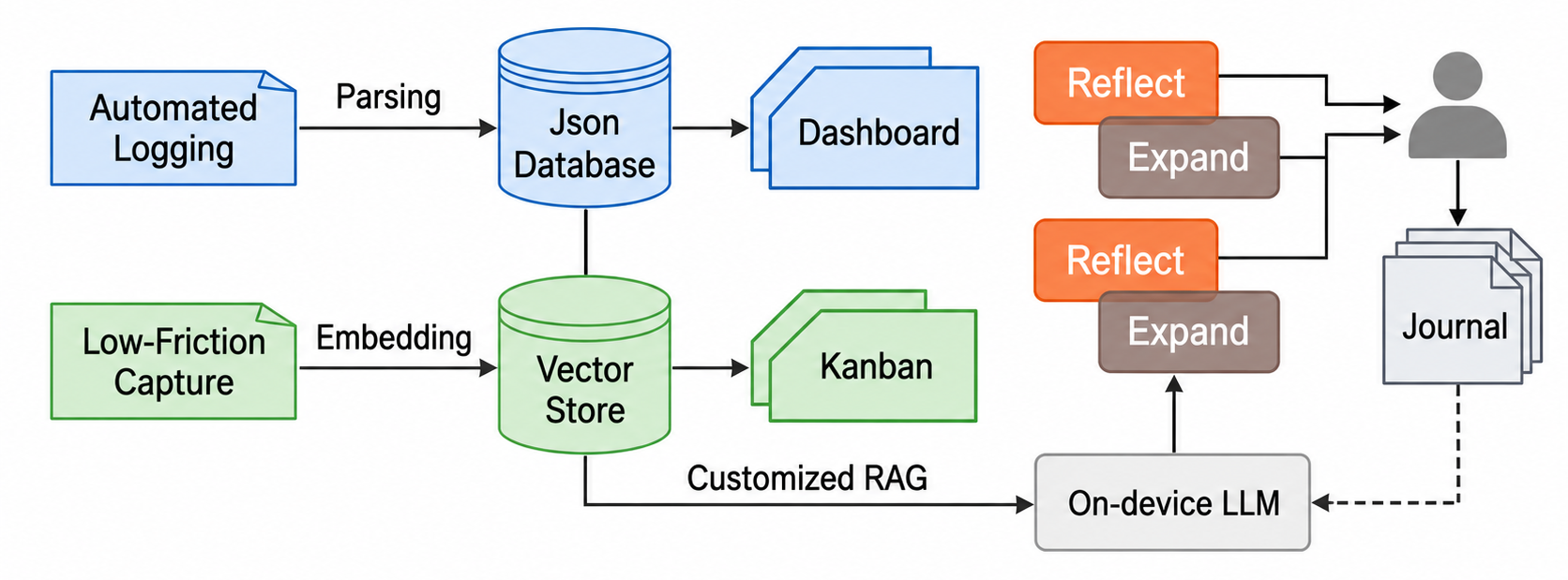}
  \caption[The generative architecture and data processing pipeline]{AuthScribe's on-device retrieval and inference pipeline. Automated logs populate the Dashboard, captured notes populate the personal context store, and retrieval combines these sources with prior Journal entries. An on-device LLM generates Reflect and Expand outputs. During the study, all retrieval and inference ran on participants' devices.}
  \Description{Pipeline diagram showing automated logs feeding an on-device structured store and Dashboard, captured notes and journals feeding an on-device vector store, retrieval combining these sources, and an on-device language model generating Reflect and Expand outputs for the writer.}
  \label{fig:authscribe-architecture}
\end{figure*}
%
%

\subsection{Interaction Workflow}

Figure~\ref{fig:authscribe-interface} shows the main interface and interaction flow. The left side provides two forms of contextual material. The Dashboard summarizes behavioral and physiological traces collected during the day and can surface patterns such as the relationship between late-night computer activity and sleep duration. The Kanban panel contains short activity or thought snippets that writers capture through a global shortcut during the day. These sources give writers concrete material they can revisit when starting or developing an entry.

The main editor contains the continuous Reflective Stream. A session begins with the writer entering an initial reflection in their own words. The writer can continue independently or request AI assistance at any point. \textit{Reflect} generates a question that helps examine an existing thought. \textit{Expand} offers a possible continuation, interpretation, or next step. Before either action, the writer selects an Interaction Stance that specifies the perspective of the response.

AI interventions appear within the chronological journal stream while retaining a visible status distinct from self-authored text. Writers can move between their own writing, Reflect questions, and Expand suggestions as the reflection develops. AuthScribe combines automatically logged traces with writer-authored Kanban snippets and prior journal entries, allowing grounding to draw on both observed behavior and experiences that writers chose to record.

\subsection{Steering and Autobiographical Grounding}

Reflect and Expand support different reflective moves. Writers use Reflect when they want to examine an experience or interpretation more closely. AuthScribe uses the current writing as a retrieval query, retrieves relevant personal context, and generates a short Socratic question. Writers use Expand when they want a possible way to develop an interpretation, response, or next step.

Interaction Stances let writers specify the perspective of each request. Stances such as \textit{Critically}, \textit{Emotionally}, and \textit{Practically} become part of the generation instruction. The writer chooses both the moment of AI intervention and the perspective through which the system responds.

Autobiographical Citations make grounding in prior self-authored material visible. When a Reflect response draws on a relevant prior journal entry or Kanban note, AuthScribe links the source inside the generated question. Hovering over the Citation opens a preview of the earlier material. Writers can inspect the connection between the present reflection and the retrieved experience and decide how that history relates to their current thinking.

\subsection{Provisional Drafts and Adoption Friction}

AuthScribe gives generated continuations a provisional status. When a writer requests Expand, the system displays the generated response as a gray, italicized, read-only Provisional Draft. Editable journal text remains the space where writers formulate their own account.

Writers can take an idea from a Provisional Draft, change its wording, make it more specific, combine it with an existing thought, or develop a different response. The interface makes the transition from suggestion to journal entry an explicit act of appropriation.

We call the effort placed at this transition \emph{adoption friction}. Interaction Stances shape what the model proposes during generation. Provisional Drafts shape how writers engage with that proposal during uptake. The design places a deliberate interaction boundary at the point where generated meaning can become part of a personally authored reflective account.

\begin{figure*}[t]
    \centering
    \includegraphics[width=0.85\textwidth]{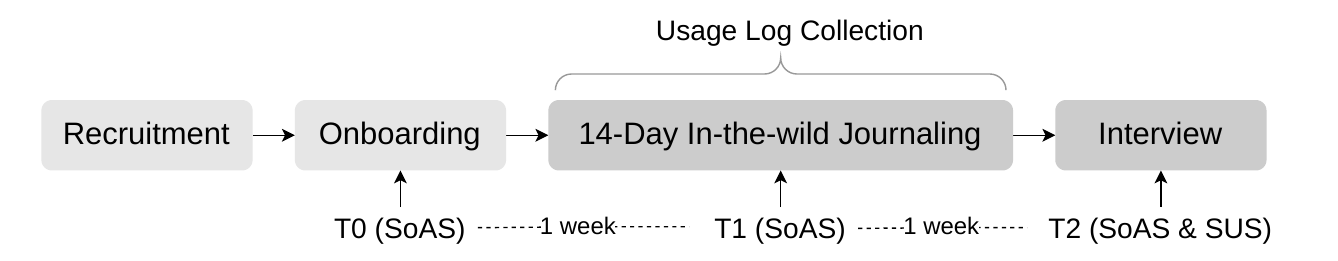}
    \caption[Overview of the 14-day field deployment study design]{Study procedure for the 14-day field deployment. Participants completed the Sense of Agency Scale at baseline (T0, Day 1), Week 1 (T1, Day 7), and Week 2 (T2, Day 14). At T2, they also completed the System Usability Scale and a semi-structured exit interview. AuthScribe collected non-semantic interaction metadata throughout the deployment.}
    \Description{Study timeline showing recruitment, onboarding, 14 days of in-the-wild journaling with continuous usage logging, and an exit interview. Participants completed SoAS at T0, T1, and T2, and completed SUS at T2.}
    \label{fig:procedure}
\end{figure*}

\subsection{On-Device Retrieval and Inference}

AuthScribe combines an Obsidian front end with an on-device retrieval and inference pipeline. The system stores structured behavioral records on-device and indexes journal entries and captured notes in an on-device vector store. When a writer invokes an AI action, AuthScribe embeds the current journal text as a query and retrieves the top-$k$ semantically relevant items from the available personal context. The study configuration used $k=3$.

AuthScribe constructs the generation input from the writer's current text, retrieved personal context, selected Interaction Stance, and the instruction for Reflect or Expand. Reflect requests a Socratic question, while Expand requests a provisional continuation. The prompt also specifies the format used for Autobiographical Citations.

During the deployment, each participant's device ran Qwen-4B through Ollama. AuthScribe also performed retrieval on-device and used Microsoft Presidio as an on-device preprocessing filter for explicit personally identifiable information. Journal entries, retrieval operations, and LLM inference stayed on the participant's computer throughout the study configuration.

Figure~\ref{fig:authscribe-architecture} summarizes the data flow from contextual sources and prior journal entries to retrieval, generation, and the corresponding interface elements.

\section{Field Deployment}
\label{sec:field-deployment}

We conducted a 14-day in-the-wild deployment to examine how people experienced AuthScribe as part of everyday reflective journaling and how these experiences evolved as personal context accumulated. The study combined longitudinal interaction metadata, repeated measures of perceived agency, an end-of-study usability measure, and semi-structured exit interviews. Figure~\ref{fig:procedure} summarizes the study timeline.


\subsection{Participants}

We recruited 14 university students, including late-stage undergraduates and postgraduate students, using purposive sampling. Participants were recruited through campus-wide announcements and word-of-mouth referrals within the student community. Eligible participants were current undergraduate or postgraduate students at our institution and had access to a personal computer suitable for the on-device deployment. Participation was voluntary and uncompensated. Recruitment materials emphasized hands-on experience with an LLM-assisted reflective journaling system as the main participation benefit.

This cohort allowed us to examine reflective journaling during a period often characterized by academic pressure, major project deadlines, and transitions toward postgraduate study or professional work \cite{Bernal_Marcos_2023,Qi_2019}. Participants ranged from 19 to 29 years of age ($M=22.21$, $SD=2.75$) and included seven men, six women, and one non-binary participant. Their academic backgrounds included computer science, engineering, business, information systems, psychology, physics, and architecture. All participants reported medium or high familiarity with generative AI. Five journaled regularly, three occasionally, three rarely, and three were first-time journalers. Table~\ref{tab:participants} summarizes the participant profiles.

\begin{table*}[t]
\centering
\caption{Participant demographics and background profiles ($N=14$).}
\label{tab:participants}
\small
\begin{tabular*}{\textwidth}{@{\extracolsep{\fill}}lllll@{}}
\toprule
ID & Age/Gender & Major & AI Familiarity & Journaling \\
\midrule
P1  & 27/F  & Information Systems & High   & Rarely \\
P2  & 20/M  & Engineering         & Medium & First-time \\
P3  & 22/F  & Business            & High   & Occasionally \\
P4  & 19/F  & Psychology          & Medium & First-time \\
P5  & 22/M  & Computer Science    & High   & Regularly \\
P6  & 21/M  & Computer Science    & High   & Occasionally \\
P7  & 20/M  & Business            & High   & Occasionally \\
P8  & 24/F  & Engineering         & High   & Regularly \\
P9  & 21/M  & Engineering         & Medium & Regularly \\
P10 & 22/F  & Computer Science    & High   & Rarely \\
P11 & 21/M  & Physics             & Medium & Regularly \\
P12 & 29/F  & Architecture        & Medium & Regularly \\
P13 & 21/NB & Computer Science    & Medium & Rarely \\
P14 & 22/M  & Computer Science    & High   & First-time \\
\bottomrule
\end{tabular*}
\end{table*}

\subsection{Procedure}

Each participant completed a 45-minute individual onboarding session before the field deployment. The session covered the study purpose, installation, and a structured walkthrough of AuthScribe. Participants installed the Obsidian plugin and configured the on-device inference environment on their own computers. The onboarding also included an offline verification step: participants temporarily disabled network connectivity and generated a test response in AuthScribe. This demonstration made local processing directly observable during setup.

The interface tutorial used dummy data to demonstrate the Dashboard, Kanban panel, Reflect and Expand actions, Interaction Stances, Autobiographical Citations, and Provisional Drafts. Participants then used AuthScribe within their normal daily routines for 14 days. They selected their own journaling topics, timing, and entry length. This deployment structure allowed reflective practices and AI use patterns to develop within participants' everyday contexts.

Participants completed the Sense of Agency Scale at three time points: baseline on Day~1 (T0), the end of Week~1 on Day~7 (T1), and the end of Week~2 on Day~14 (T2). At T2, participants also completed the System Usability Scale and a semi-structured exit interview. Interviews lasted 40--60 minutes and took place either online through Zoom or in person according to participant preference.

Our institutional human research ethics committee approved the study. Participation was voluntary, and participants could skip journal entries, set aside AI-generated suggestions, or withdraw from the study. The onboarding framed AuthScribe as a personal knowledge management and self-reflection tool. Participants also received relevant university and emergency support resources as part of the study safeguards.

\subsection{Data Collection}

AuthScribe collected non-semantic interaction metadata throughout the 14-day deployment. Logged fields included timestamps, session duration, word counts, Interaction Stance selections, AI action types, and the source category used by the retrieval pipeline, including Dashboard, Kanban, and Journal context. These logs supported analyses of engagement patterns, interaction sequences, and changes in contextual grounding over time.

Journal text and personal Kanban content remained on each participant's device throughout the study. The researcher-facing dataset contained interaction metadata, survey responses, and exit-interview transcripts. This data-collection strategy preserved access to behavioral and experiential evidence while keeping the semantic contents of participants' journals on their own devices.

The exit interviews explored participants' expectations, experiences with contextual anchors and Autobiographical Citations, use of Reflect and Expand, perceptions of Provisional Drafts, authorship and ownership, privacy, and overall usability. The interview guide invited concrete examples from the two-week deployment, including wording that participants recalled from specific AI interactions and their own re-expressions. We treated these recollections as interview data, while the underlying journal corpus remained on participants' devices. All qualitative accounts of change across the deployment came from these end-of-study retrospective interviews.

\subsection{Measures}

\paragraph{System Usability Scale.}
At T2, participants completed the 10-item System Usability Scale (SUS) \cite{brooke1996sus,brooke2013sus}. Each item uses a five-point Likert response, and the standard scoring procedure produces a value from 0 to 100. We use SUS to characterize the overall usability of the integrated AuthScribe experience. Feature-specific experiences, including the effort associated with Provisional Drafts, are examined through the interview data.

\paragraph{Sense of Agency Scale.}
Participants completed the Sense of Agency Scale (SoAS) at T0, T1, and T2 \cite{tapal2017sense}. We used the 11-item factor structure reported in the original scale's confirmatory analysis, comprising five Sense of Positive Agency (SoPA) items and six Sense of Negative Agency (SoNA) items. Responses used a seven-point Likert scale.

SoAS provides a complementary longitudinal measure of participants' general perceived agency during the deployment. We interpret authorship and ownership primarily through participants' accounts of their interactions with generated language, while SoAS characterizes broader changes in perceived control across the study period.

\subsection{Analysis}

\paragraph{Interaction metadata.}
We summarized system use across participants and time, including counts of self-authored entries, Reflect and Expand events, word-count patterns, and retrieval-source distributions. We compared descriptive patterns between Week~1 and Week~2 and examined day-level changes in the use of Dashboard, Kanban, and Journal context. We also summarized co-occurrences between Interaction Stances and retrieval-source categories. These analyses characterize observable interaction patterns and provide behavioral context for the survey and interview findings.

\paragraph{Longitudinal agency.}
We analyzed SoPA and SoNA scores using separate one-way repeated-measures ANOVAs with time (T0, T1, T2) as the within-participant factor \cite{field2024discovering}. Mauchly's tests indicated that the sphericity assumption was satisfied for both models. We report the standard degrees of freedom ($df_{\mathrm{time}}=2$, $df_{\mathrm{error}}=26$). Significant omnibus effects were followed by Bonferroni-adjusted pairwise comparisons across T0, T1, and T2. We report partial eta squared ($\eta_p^2$) as the effect-size measure. These analyses characterize temporal patterns within the study cohort.

\paragraph{Interview analysis.}
We analyzed the exit interviews using reflexive thematic analysis (RTA) \cite{braun2006using,braun2019reflecting}. The interview guide foregrounded study-relevant concepts including contextual grounding, authorship, privacy, and usability, while participants' examples provided additional patterns grounded in their own experiences. We used an abductive analytic orientation, moving iteratively between these sensitizing concepts and patterns developed through repeated reading of the transcripts.

One researcher conducted transcript familiarization, initial coding, theme construction, and iterative theme refinement. The researcher's familiarity with the study design and AuthScribe informed the reflexive interpretation of participants' accounts. The analysis emphasized relationships across participants' experiences, including retrospective descriptions of change in their responses to AI-generated suggestions and autobiographical grounding. An audit trail linked developing themes to supporting excerpts and analytic notes.

\section{Findings}
\label{sec:findings}

Participants rated AuthScribe's overall usability highly at the end of the deployment ($M=81.5$, $SD=6.4$). Interaction logs locate changes in participants' use across the two weeks, and the interviews explain how participants experienced those changes. Four findings organize this section: grounding changed the evidence participants used to interpret themselves; perceived agency shifted after participants had established a practice of use; authorship emerged through transformation and commitment to AI-generated suggestions; and autobiographical provenance strengthened trust in personalized assistance.


\subsection{Grounding Changed What Counted as Evidence for Reflection}
\label{sec:findings-context}

Participants initiated reflection at a similar rate across the deployment. They created 81 self-authored starting entries in Week~1 and 84 in Week~2. Reflect events increased from 45 to 113, and Expand events increased from 51 to 90 (Table~\ref{tab:usage}). The main longitudinal change appeared after participants had already put an experience into their own words. Participants increasingly used AI to develop, question, and reconnect those initial reflections.

\begin{table*}[t]
\centering
\caption{System usage across the two study weeks. Word values report $M{\pm}SD$ per user-authored entry or AI-generated block.}
\label{tab:usage}
\small
\begin{tabular*}{\textwidth}{@{\extracolsep{\fill}}lrrrrrr@{}}
\toprule
& \multicolumn{2}{c}{Initial}
& \multicolumn{2}{c}{Reflect}
& \multicolumn{2}{c}{Expand} \\
\cmidrule(lr){2-3}
\cmidrule(lr){4-5}
\cmidrule(lr){6-7}
Week
& Count & Words/entry
& Count & Words/block
& Count & Words/block \\
\midrule
Week 1
& 81 & $38.2{\pm}7.5$
& 45 & $49.4{\pm}8.8$
& 51 & $102.4{\pm}16.6$ \\
Week 2
& 84 & $37.8{\pm}8.6$
& 113 & $86.5{\pm}14.6$
& 90 & $125.7{\pm}17.8$ \\
\bottomrule
\end{tabular*}
\end{table*}

Early in the deployment, contextual panels helped participants turn diffuse experience into a specific episode they could examine. P4 described a sleep graph beside screen-time data as ``a concrete starting point.'' P10 described captured fragments as material ready to assemble into a reflection. These traces gave participants an evidentiary foothold: a concrete event, behavior, or moment that could anchor interpretation.

The scale of interpretation also changed. P5 described stress as coloring an entire week as a failure. Dashboard traces localized that judgment to a particular period and shifted the participant's question from ``why is my life falling apart?'' to ``what exactly triggered this shift on Thursday afternoon?'' P8 connected a diffuse feeling of distress to a specific lab meeting through Kanban notes. Factual context converted broad self-judgments into situated questions about specific episodes. Participants could examine circumstances, triggers, and sequences of events as evidence for understanding how they felt.

\begin{figure*}[t]
    \centering
    \includegraphics[width=0.90\linewidth]{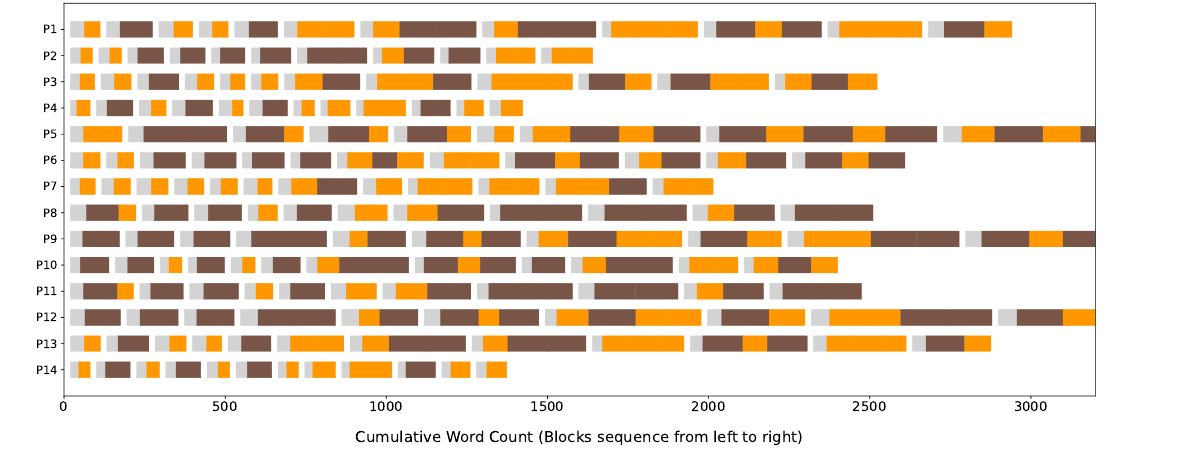}
    \caption{Reflect and Expand use across the 14-day deployment. Block width represents word count, and block order follows each participant's reflective stream.}
    \label{fig:narrative-sequences}
\end{figure*}

Figure~\ref{fig:narrative-sequences} shows that participants developed different rhythms for carrying these reflections forward. Some alternated between Reflect and Expand, while others used one action more heavily. The common structure remained participant-initiated: a self-authored entry established the experience, and AI actions became resources for developing it.

\begin{figure}[H]
    \centering
    \includegraphics[width=\linewidth]{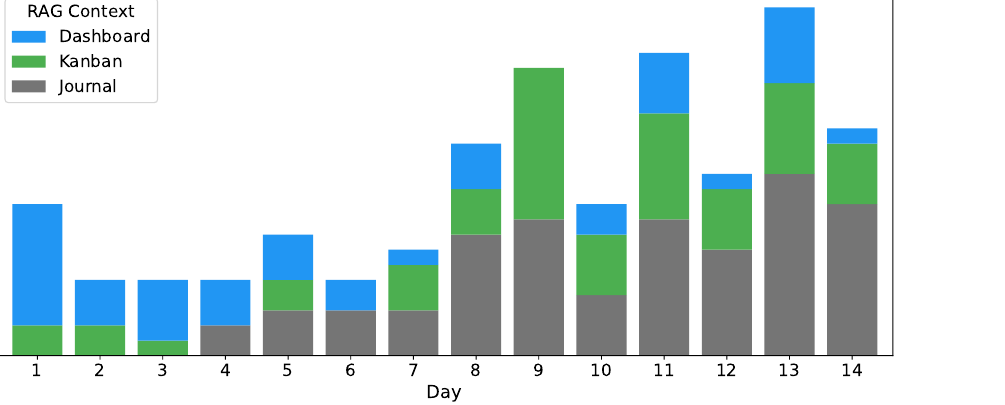}
    \caption{Retrieved context across the 14-day deployment. Dashboard use decreased as Kanban and Journal material appeared more often during Reflect events.}
    \label{fig:rag-over-time}
\end{figure}

Participants later accumulated enough writing to extend reflection across time. Dashboard retrieval decreased across the two weeks, while Journal and Kanban retrieval increased (Table~\ref{tab:rag-sources}; Figure~\ref{fig:rag-over-time}). Early grounding established recent circumstances. Later grounding resurfaced interpretations participants had already written about related experiences. Current experiences could then sit beside prior self-authored meanings.

\begin{table}[H]
\centering
\caption{Average daily retrieval-source invocations per participant.}
\label{tab:rag-sources}
\begin{tabular*}{\linewidth}{@{\extracolsep{\fill}}lccc@{}}
\toprule
Week & Dashboard & Kanban & Journal \\
\midrule
Week 1 & $0.55{\pm}0.55$ & $0.23{\pm}0.42$ & $0.25{\pm}0.44$ \\
Week 2 & $0.23{\pm}0.49$ & $0.54{\pm}0.70$ & $0.84{\pm}0.71$ \\
\bottomrule
\end{tabular*}
\end{table}

Behavioral traces and autobiographical writing supported different reflective moves. Behavioral traces helped participants ask \emph{what happened here?} Accumulated journal entries helped them ask \emph{how does this connect to what I have understood about myself before?} The growing journal record changed the evidence available for reflection. Participants moved from examining a current episode in context to interpreting the present through meanings they had authored across earlier sessions.

This change also shifted the role of AI grounding. Early context helped participants narrow an experience to something they could investigate. Later context could return an earlier interpretation and place it beside the present one. The system increasingly mediated a relationship between the participant's current account and their earlier accounts. The longitudinal value of personalization came from this expanding self-authored record, which let participants compare experiences across time and revisit their own prior reasoning.

\begin{figure}[H]
    \centering
    \includegraphics[width=\linewidth]{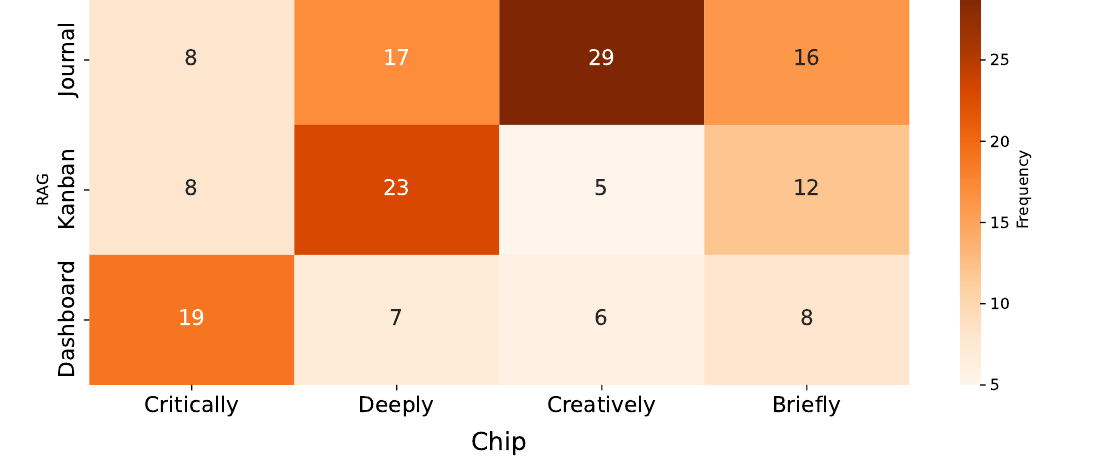}
    \caption{Co-occurrence between selected Interaction Stances and retrieved context sources during Reflect events.}
    \label{fig:stance-source}
\end{figure}

Figure~\ref{fig:stance-source} offers a secondary descriptive view of stance and retrieval source. Dashboard-grounded Reflect events most often co-occurred with \textit{Critically} (47.5\%, $n=19$), Kanban-grounded events with \textit{Deeply} (47.9\%, $n=23$), and Journal-grounded events with \textit{Creatively} (41.4\%, $n=29$). Retrieval followed semantic similarity to the current text, and participants selected stance explicitly. These co-occurrences describe the combinations of personal material and inquiry perspectives that appeared during use.

\subsection{Perceived Agency Shifted After Participants Established a Practice of Use}
\label{sec:findings-agency}

Participants' Sense of Agency scores followed a two-phase trajectory. SoPA stayed close to baseline at Week~1 and increased at Week~2. SoNA stayed close to baseline at Week~1 and decreased at Week~2. Time had a significant effect on SoPA, $F(2,26)=11.69$, $p<.001$, $\eta_p^2=.47$, and SoNA, $F(2,26)=13.35$, $p<.001$, $\eta_p^2=.51$ (Table~\ref{tab:agency-anova}).

\begin{table}[H]
\centering
\caption{Sense of Agency across the three study time points ($N=14$).}
\label{tab:agency-anova}
\begin{tabular*}{\linewidth}{@{\extracolsep{\fill}}lccccc@{}}
\toprule
Dimension & T0 & T1 & T2 & $F(2,26)$ & $\eta_p^2$ \\
\midrule
SoPA & $4.29{\pm}0.35$ & $4.37{\pm}0.47$ & $4.85{\pm}0.78$ & 11.69 & .47 \\
SoNA & $4.49{\pm}0.52$ & $4.44{\pm}0.45$ & $3.98{\pm}0.65$ & 13.35 & .51 \\
\bottomrule
\end{tabular*}
\end{table}

The pairwise comparisons locate the change in the second week. T0 and T1 stayed statistically similar for both dimensions. T1 and T2 differed for SoPA ($p=.013$) and SoNA ($p=.006$), and T0 and T2 also differed for both dimensions (Table~\ref{tab:agency-posthoc}). The timing matters because AuthScribe offered the same interaction controls from Day~1. Participants' broad perceived agency changed after those controls had become part of repeated journaling sessions.

\begin{table}[H]
\centering
\caption{Bonferroni-adjusted pairwise comparisons for Sense of Agency.}
\label{tab:agency-posthoc}
\begin{tabular*}{\linewidth}{@{\extracolsep{\fill}}lccc@{}}
\toprule
& \multicolumn{3}{c}{Adjusted $p$} \\
\cmidrule(lr){2-4}
Dimension & T0 vs.\ T1 & T1 vs.\ T2 & T0 vs.\ T2 \\
\midrule
SoPA & .900  & \textbf{.013} & \textbf{.020} \\
SoNA & 1.000 & \textbf{.006} & \textbf{.017} \\
\bottomrule
\end{tabular*}
\end{table}

By Week~2, participants had repeatedly decided when to invoke AI, how to steer it, how to transform generated language, and how to interpret prompts grounded in growing personal context. The interface had offered these opportunities throughout the deployment; sustained use gave participants repeated occasions to exercise them. The SoAS trajectory places the agency shift within this developing practice.

This separation between available controls and perceived agency is central to the longitudinal result. A control can exist from the first interaction, while its role in a person's practice develops through repeated decisions. Participants learned where AI assistance fit into their journaling, when a suggestion deserved further work, and how much of their own history they wanted to bring into a current reflection. Their broader sense of agency changed after participants had accumulated repeated experience with these interactions.

SoAS measures broad perceived agency across the deployment. Interview accounts of rewriting provide more direct evidence about authorship and ownership. The two forms of evidence operate at different levels: the scale captures a longitudinal shift in perceived control, while the interviews show how participants enacted control through concrete decisions about steering, interpretation, and uptake.

\subsection{Authorship Emerged Through Transformation and Commitment}
\label{sec:findings-rewriting}

Participants located authorship in what they did with AI-generated suggestions. Early in the deployment, the uptake boundary could feel costly. P2 recalled receiving a useful resolution late at night and wanting to ``click `accept' and move on to sleep.'' Speed mattered in that moment.

Eleven of 14 participants later described value in rewriting. P1's example shows the mechanism clearly. The AI proposed ``I must fortify my boundaries,'' and the participant entered ``I need to stop saying yes to every meeting.'' The transformation changed register, specificity, and personal commitment. An abstract proposal became a situated statement about the participant's own behavior.

Re-expression made differences between generated language and personally acceptable expression actionable. Participants evaluated a candidate interpretation in their own circumstances and transformed it into language they chose to place in the journal. The generated suggestion supplied possible meaning; the writer decided how that meaning applied to their life and which formulation they were willing to endorse. Authorship emerged through interpretation, selection, transformation, and commitment.

P9 captured the process succinctly: ``By the time I finished typing, the thought was mine.'' The initial idea could originate in an AI suggestion, and ownership emerged through decisions about meaning, expression, and inclusion in the participant's account. P6 described the read-only block as a clear boundary between the provisional source and the journal entry. These accounts locate authorship in appropriation: participants could accept the usefulness of an AI-generated idea while still performing the work that gave the idea personal meaning and final form.

The distinction matters in reflective journaling because a sentence can carry an interpretation of the writer's own experience. Fluency makes an interpretation easy to receive. Re-expression asks the writer to decide whether that interpretation deserves endorsement and how it should appear in their personal account. The interaction turned uptake into a moment of judgment about meaning as well as wording.

The boundary also carried effort. Fatigue increased the value of immediate completion, and rewriting required work. Participants still described value in the transformation. The Provisional Draft became a site where they negotiated convenience, interpretation, and responsibility for the words that entered their journal.

\subsection{Personalization Became Trustworthy When Participants Could Trace Its Basis}
\label{sec:findings-trust}

Participants trusted personalized assistance when they could understand why AuthScribe related a prompt to them. Ten of 14 participants described Autobiographical Citations as contributing to relevance or trust, and their explanations pointed to recognizable provenance in their own prior writing.

P3 characterized early Dashboard-grounded interaction as ``a strict data analyst just pointing out my screen time.'' A later prompt cited an earlier entry about a similar insecurity, and P3 described the system as feeling like it ``actually knew me.'' Behavioral traces connected the prompt to recent activity. An Autobiographical Citation connected it to an interpretation that P3 had written earlier. The journal source gave P3 recognizable evidence for the system's apparent knowledge.

P7 described the basis of trust directly: ``Seeing my own words hyperlinked as a reference right above the AI's question made me trust the prompt. It was a continuation of my own thought process from last Tuesday.'' The citation let P7 trace a current question to prior reasoning. P12 described a related experience as a reminder of a lesson from an earlier struggle. Personalized assistance gained credibility through continuity with participants' own prior thinking.

Autobiographical Citations created a visible evidentiary chain: a current concern connected to a retrieved past entry, and that entry supported the generated question. Participants could inspect each connection and judge whether the system's use of their history made sense. Personalization gained credibility through this traceable relationship to self-authored evidence.

The source of credibility is important. The AI could surface an earlier interpretation, commitment, or lesson, while the participant recognized that material as something they had previously written. The system's apparent understanding drew authority from the participant's own past reasoning. AuthScribe acted as a mediator between moments in the participant's reflective history and returned earlier self-interpretations to the present for reconsideration.

A smaller group described traceability around data processing. During onboarding, participants disabled network connectivity and observed AuthScribe continue to generate a response. P6 connected this verification to greater comfort with intimate disclosure. P1 associated the local processing configuration with a stronger sense of private ownership over the journal space. Participants could connect the privacy claim to an observable property of the system.

Traceability shaped trust at two points in the interaction. Citations showed why a prompt related to the participant through autobiographical provenance. Offline verification showed where personal material was processed. Both mechanisms gave participants concrete evidence about how AuthScribe used their data.

Autobiographical Citations made the path from prior self-authored material to AI assistance inspectable. Provisional Drafts made the path from AI assistance to personal writing deliberate. Participants could examine the basis of personalized assistance and decide how AI-generated suggestions became personally accountable language. These two transitions ground the design implications in the next section.

\section{Discussion}
\label{sec:discussion}

Participants encountered two consequential transitions in LLM-assisted reflective journaling. Personal context entered AI assistance through retrieval. Autobiographical Citations exposed the prior self-authored material used to ground prompts. AI-generated suggestions entered personal writing through Provisional Drafts and re-expression. Over time, participants learned how to judge what AI-generated suggestions meant for them, transform those suggestions into personally accountable language, and evaluate why personalized prompts related to their own history. We discuss three implications for AI systems that support personal reflection: authorship as appropriation at the uptake boundary, agency as a practice of use, and personalization as an accountable relationship to autobiographical evidence.

\subsection{Adoption Friction Locates Authorship at the Uptake Boundary}
\label{sec:discussion-adoption-friction}

Human--AI writing research increasingly treats authorship as an interactional outcome shaped by how writers evaluate, revise, and develop AI-generated suggestions \cite{Draxler_2024,Reza_2025,kadoma2024ownership,polimetla2026ownership,bhat2026reactive}. AI-generated suggestions can seed ideas and interpretations that writers later elaborate, shifting part of the writing process toward judgment, verification, and integration \cite{jakesch2023opinionated,bhat2026reactive,lee2025critical}. In reflective journaling, this shift reaches the interpretation of personal experience itself. A generated continuation can offer a plausible account of what an event means before the writer has fully articulated that meaning.

Our findings locate authorship in the work participants performed after receiving a candidate interpretation. Re-expression changed vocabulary, specificity, framing, and personal commitment. A generic suggestion could become a situated statement about the participant's own behavior, priorities, or experience. Appropriation captured the key action: participants interpreted a generated idea in relation to their own circumstances and chose the form they wanted to claim. The origin of a candidate interpretation and the authorship of the final reflective statement became separate stages. Transformation, selection, and commitment let participants take ownership of AI-generated suggestions.

Recent HCI work emphasizes meaningful human judgment during AI-supported writing and shows how carefully placed interaction effort can encourage deliberation \cite{nishal2026agency,carrera2026egg,andersson2026tombwriter,arun2026friction}. We use \emph{adoption friction} to name effort placed at the transition where an AI-generated suggestion acquires status inside a human-authored artifact. In AuthScribe, visual separation marked AI output as provisional, and re-expression made uptake an active decision. The design contribution centers on this location of friction and the appropriation work that participants performed there.

Interaction Stances and Provisional Drafts supported different forms of control. Interaction Stances shaped the direction of generation. Provisional Drafts shaped the status of AI-generated suggestions after they appeared. Reflective journaling systems increasingly support user-directed exploration and interpretive feedback \cite{diarymate,song2025exploreself,hwang2026augmentiary}. Our findings extend control beyond generation to the decisions that turn a suggestion into personally accountable language.

The boundary also imposed effort. Participants described moments where fatigue increased the value of immediate completion, and they later described value in the transformation required by rewriting. Future designs need to calibrate this effort to the artifact and the moment of use. Writers may value stronger uptake friction when an artifact carries identity, responsibility, or personal commitment. They may value lighter interaction when the immediate goal emphasizes capture or momentum. Structured paraphrasing, staged acceptance, selective re-expression, and user-adjustable transformation prompts offer several ways to vary the strength of the uptake boundary.

\subsection{Agency as a Practice of Use}
\label{sec:discussion-agency}

Interfaces often instantiate control through visible choices: users request assistance, select a direction, edit generated text, or inspect system output \cite{amershi2019guidelines,wu2022aichains,pan2024new,Reza_2025,zhou2026legoui}. AuthScribe offered these choices from the first day of the deployment. Participants decided when to invoke AI, chose an Interaction Stance, inspected Autobiographical Citations, and responded to Provisional Drafts.

The longitudinal agency pattern adds a temporal dimension to this view of control. Participants' SoAS scores stayed close to baseline through Week~1 and shifted during Week~2. By then, participants had used the same controls across repeated journaling sessions, invoked AI assistance more frequently, gained familiarity with rewriting, and accumulated more personal context. The availability of control and the experience of agency unfolded on different timescales.

Recent studies of AI-assisted writing show that self-efficacy, trust, ownership, prompting strategies, and expectations of AI participation can change across repeated interaction \cite{park2026authorship,qin2026personalization,nishal2026agency,rismani2026oversight,zhang2026deployability}. Our findings extend this temporal perspective to reflective journaling. Participants developed practical knowledge about when AI support helped, how they wanted it steered, which AI-generated suggestions deserved more work, and how prior self-authored material should inform a current reflection. Participants exercised agency through these recurring decisions as their broader SoAS scores shifted over the same period.

We treat agency in longitudinal AI systems as a \emph{practice of use}. Interface affordances create opportunities for control. Repeated interaction gives people experience in exercising, interpreting, and adapting those opportunities. A single session can show whether users understand and operate a control. Longitudinal use can show how that control gains meaning within an evolving personal practice and how its role changes as the user and the system accumulate history.

The personal context available to AuthScribe changes through use. Participants created new journal entries, revisited previous interpretations through Citations, and developed habits around Reflect, Expand, and rewriting. The second-week agency shift occurred within this broader process of adaptation. Several mechanisms developed during the same period, so future evaluations should examine how user practices and available personal context co-evolve over time. Staged introductions, comparative uptake conditions, and controlled variation in available history can help isolate specific mechanisms while preserving longitudinal use.

\subsection{Personalization Becomes Accountable Through Autobiographical Provenance}
\label{sec:discussion-provenance}

AI systems for personal reflection increasingly draw on behavioral traces, prior writing, and autobiographical memories to generate context-sensitive support \cite{nepal2024mindscape,zulfikar2025resonance,ozawaburns2025remigraph,hwang2026augmentiary}. This literature shows the value of connecting reflection to lived experience. Our findings add a user-centered question about the basis of that connection: participants cared about whether they could recognize why a prompt related to them.

The shift from Dashboard grounding toward Journal and Kanban grounding changed the evidence participants encountered. Behavioral traces helped them localize diffuse self-judgments into specific episodes. Accumulated journal material resurfaced interpretations they had previously authored and connected present reflection with earlier meanings. Participants encountered different forms of evidence about themselves, and each form supported a different reflective move.

Autobiographical Citations made this evidence inspectable. Research on human--AI reliance and transparent interaction shows that visible sources and understandable system structure can help people evaluate AI output and exercise oversight \cite{kim2025reliance,wu2022aichains,shanmugarasa2025privacy,rismani2026oversight}. In AuthScribe, participants traced a current reflective question to their own prior writing and judged the relevance of that connection. We call this property \emph{accountable autobiographical provenance}: personalized assistance carries a visible relationship to the personal evidence that supports it.

This provenance gives personalized assistance a user-visible basis for credibility. Participants trusted prompts that continued a line of reasoning they recognized from earlier writing. The suggestion gained credibility through continuity with the user's prior self. AI acted as a mediator across moments in the participant's reflective history by returning earlier interpretations to the present for reconsideration. In journaling, this continuity contributes directly to meaning-making across time.

Designers can strengthen this relationship by exposing source type, temporal distance, and the specific self-authored material associated with a suggestion. Users can assess the retrieved history and decide what it means in the present. This approach gives people an active role in evaluating personalization and supports the longitudinal value of accumulated personal context.

Local processing created a related form of accountability. During onboarding, participants verified that AuthScribe continued to operate with network connectivity disabled, and some connected that observation to greater comfort with intimate disclosure. This privacy evidence came from fewer participants than the citation finding, so we treat it as a secondary pattern. The interaction principle remained the same: participants could evaluate consequential relationships to personal data when the system made those relationships observable.

Autobiographical Citations expose the path from prior self-authored material to AI assistance. Provisional Drafts expose the path from AI assistance to personal writing. These boundaries show users why the system contributes a particular suggestion and give them an active role in deciding what that suggestion becomes. AI systems for personal reflection can support authorship and trust by making these transitions inspectable, interpretable, and actionable across time.

\section{Limitations}
\label{sec:limitations}

This study examined AuthScribe with 14 university students over two weeks. The cohort captures reflective journaling during academic and career transitions. Broader populations and longer deployments can extend these findings across different journaling practices, privacy expectations, and levels of technical familiarity.

Participants used the full AuthScribe configuration throughout the deployment, while familiarity, AI use, rewriting experience, and personal context all developed across the same period. The findings characterize a system-level longitudinal trajectory. Exit interviews occurred at T2, so participants' qualitative accounts of change across the two weeks reflect retrospective recall. Future comparative deployments can vary individual mechanisms and assessment timing to examine these processes more closely.

SoAS provides a broad longitudinal measure of perceived agency, while interview accounts provide the primary evidence for authorship and ownership. The journal corpus remained on participants' devices, so semantic analysis drew on concrete examples that participants recounted during interviews. Future studies can combine task-specific agency measures with participant-controlled sharing protocols when linguistic appropriation or reflective depth becomes a primary outcome.

\section{Conclusion}
\label{sec:conclusion}

AuthScribe examines how writers incorporate AI assistance into reflective journaling while retaining an active role in the transition from suggestion to authored text. 
In a 14-day deployment with 14 participants, contextual grounding shifted from recent behavioral traces toward accumulated self-authored material, perceived agency changed during the second week, and participants described rewriting Provisional Drafts as a process of transforming and claiming AI-generated suggestions. 
Autobiographical Citations also gave participants a visible basis for judging personalized prompts.

The study develops three design ideas for AI systems that support personal reflection. 
Adoption friction places user effort at the point where an AI-generated suggestion enters a human-authored artifact. Accountable autobiographical provenance lets users inspect the personal evidence behind personalized assistance.
A longitudinal view of agency attends to how experienced control changes across repeated use. 
These ideas support AI systems for personal reflection that keep personal context and AI-generated language visible at the moments where users interpret and adopt them.

\paragraph{Generative AI Use Disclosure.}
The authors used ChatGPT (OpenAI) to revise language and reorganize existing manuscript content for clarity and concision. The authors reviewed and verified all changes and take responsibility for the final manuscript.

\bibliographystyle{ACM-Reference-Format}
\bibliography{reference}

\newpage
\appendix

\section{Semi-Structured Exit Interview Guide}
\label{app:interview-guide}

The exit interview lasted 40–60 minutes and took place at the end of the 14-day deployment. The questions below reproduce the documented interview guide, grouped by topic for readability.

\subsection{General Expectations and Baseline}

\begin{itemize}
    \item Before using this system, what were your general expectations regarding the use of an AI to assist with personal, intimate journaling?
    \item How did your actual experience over the past 14 days align with or diverge from those initial expectations?
\end{itemize}

\subsection{Contextual Anchors}

\begin{itemize}
    \item Often, users experience a ``blank page syndrome'' when trying to start a journal. How did the presence of the Dashboard (objective data) and Kanban (fleeting snippets) affect your ability to initiate a journal entry?
    \item Can you recall a specific instance where seeing a metric (e.g., sleep, gaming time) on the Dashboard changed the way you framed or understood your stress that day?
\end{itemize}

\subsection{Autobiographical Grounding and Citations}

\begin{itemize}
    \item During the second week, the system frequently utilized your past journal entries as citations for its questions. How did you feel when the AI referenced a specific thought or vulnerability you had written previously?
    \item Did this historical grounding change your perception of the AI (e.g., from a tool to a reflective partner)? Did it influence your willingness to use the Interaction Stance chips to direct the AI?
\end{itemize}

\subsection{Provisional Drafts and Authorship}

\begin{itemize}
    \item The AI's suggested drafts were deliberately designed to be read-only (gray, italicized text), preventing you from simply copying and pasting them into your journal. What was your initial reaction to this design constraint?
    \item As you continued to use the system, how did the act of manually transcribing and rewriting the AI's suggestions affect your sense of ownership over the final text? Did you find yourself changing the AI's words to better match your authentic voice?
\end{itemize}

\subsection{Privacy and Usability}

\begin{itemize}
    \item Knowing that the system ran entirely locally on your device without transmitting your journal entries to the cloud, did this absolute privacy guarantee influence your level of self-disclosure?
    \item Were there any aspects of the system's interface that felt disruptive to your natural reflective flow?
\end{itemize}

\newpage
\section{Administered Sense of Agency Scale}
\label{app:soas}

Participants completed the 11-item Sense of Agency Scale (SoAS) used in the study at T0, T1, and T2 \cite{tapal2017sense}. They rated each statement on a seven-point Likert scale from 1 (\textit{Strongly Disagree}) to 7 (\textit{Strongly Agree}). The administered form used the 11-item factor structure reported in the original scale's confirmatory analysis, with five Sense of Positive Agency (SoPA) items and six Sense of Negative Agency (SoNA) items. Original Items 4 and 5 were excluded from the administered form.

\begin{table}[!htbp]
\centering
\caption{The 11-item Sense of Agency Scale administered in the study.}
\label{tab:appendix-soas}
\small
\begin{tabularx}{\linewidth}{@{}cXc@{}}
\toprule
Item & Statement & Dimension \\
\midrule
1  & I am in full control of what I do. & SoPA \\
2  & I am just an instrument in the hands of somebody or something else. & SoNA \\
3  & My actions just happened without my intention. & SoNA \\
6  & My movements are automatic---my body simply makes them. & SoNA \\
7  & The outcomes of my actions generally surprise me. & SoNA \\
8  & Things I do are subject only to my free will. & SoPA \\
9  & The decision whether and when to act is within my hands. & SoPA \\
10 & Nothing I do is actually voluntary. & SoNA \\
11 & While I am in action, I feel like I am a remote controlled robot. & SoNA \\
12 & My behavior is planned by me from the very beginning to the very end. & SoPA \\
13 & I am completely responsible for everything that results from my actions. & SoPA \\
\bottomrule
\end{tabularx}
\end{table}

\section{AuthScribe Generation and Grounding Configuration}
\label{app:generation-config}

AuthScribe generated AI assistance from four inputs: the writer's current journal text, retrieved personal context, the selected Interaction Stance, and the requested action. The study configuration retrieved the top three semantically relevant context items ($k=3$) from the available personal context. Eligible context sources included Dashboard records, Kanban snippets, and prior Journal entries.

\subsection{Interaction Stances}

Reflect and Expand used separate stance sets in the deployed interface. Writers selected one stance before generation.

\begin{table}[!htbp]
\centering
\caption{Interaction Stances available for Reflect and Expand.}
\label{tab:appendix-stances}
\begin{tabular*}{\linewidth}{@{\extracolsep{\fill}}ll@{}}
\toprule
Action & Available Interaction Stances \\
\midrule
Reflect & Deeply, Critically, Briefly, Creatively \\
Expand & Logically, Emotionally, Practically, Abstractly \\
\bottomrule
\end{tabular*}
\end{table}

\subsection{Retrieval and Generation Structure}

For each AI request, AuthScribe embedded the current journal text as the retrieval query and selected the top-$k$ personal-context items by semantic similarity. The system then constructed a generation request from the current text, retrieved context, selected stance, and action instruction.

Reflect requested a short Socratic question that examined the writer's current reflection. When retrieved Journal or Kanban material informed the question, the generation instruction supplied Obsidian-compatible link formatting so the response could render Autobiographical Citations with hover previews.

Expand requested a provisional continuation, interpretation, or possible next step. AuthScribe displayed this output as a read-only Provisional Draft in the Reflective Stream. The writer could use the draft as material for re-expression in the editable journal.

\begin{table}[!htbp]
\centering
\caption{Inputs and output behavior for the two AI actions.}
\label{tab:appendix-generation}
\small
\begin{tabularx}{\linewidth}{@{}lXX@{}}
\toprule
 & Reflect & Expand \\
\midrule
Current text & Used as retrieval query and generation context & Used as retrieval query and generation context \\
Personal context & Top-$k$ relevant Dashboard, Kanban, or Journal material & Top-$k$ relevant Dashboard, Kanban, or Journal material \\
User steering & One Reflect Interaction Stance & One Expand Interaction Stance \\
Generation objective & Short Socratic question for further examination & Provisional continuation, interpretation, or next step \\
Interface output & Reflect response with Autobiographical Citations when grounded in retrieved personal writing & Read-only Provisional Draft in the Reflective Stream \\
\bottomrule
\end{tabularx}
\end{table}

During the deployment, each participant's device ran Qwen-4B through Ollama, and AuthScribe performed retrieval on-device, as described in Section~3.5.

\end{document}